# Delay Tolerant Control for Autonomous Driving Using CDOB

Xincheng Cao, Haochong Chen, Levent Guvenc and Bilin Aksun-Guvenc

Automated Driving Lab, Ohio State University

**Abstract**

With the rapid growth of autonomous vehicle technologies, effective path-tracking control has become a critical component in ensuring safety and efficiency in complex traffic scenarios. When a high-level decision-making agent generates a collision-free path, a robust low-level controller is required to precisely follow this trajectory. However, connected autonomous vehicles (CAV) are inherently affected by communication delays and computation delays, which significantly degrade the performance of conventional controllers such as PID or other more advanced controllers like disturbance observers (DOB). While DOB-based designs have shown effectiveness in rejecting disturbances under nominal conditions, their performance deteriorates considerably in the presence of unknown time delays. To address this challenge, this paper proposes a delay-tolerant communication disturbance observer (CDOB) framework for path-tracking control in delayed systems. The proposed CDOB compensates for the adverse effects of time delays, maintaining accurate trajectory tracking even under uncertain and varying delay conditions. It is shown through a simulation study that the proposed control architecture maintains close alignment with the reference trajectory across various scenarios, including single-lane change, double-lane change, and Elastic Band–generated collision-avoidance paths under various time delays. Simulation results further demonstrate that the proposed method outperforms conventional approaches in both tracking accuracy and delay robustness, making it well-suited for connected autonomous driving applications.

**Keywords:** delay tolerant control; autonomous vehicle control; communication disturbance observer; path tracking control

# 1. Introduction

With rapid urbanization and technological development, the number of privately owned vehicles has dramatically increased each year. The excessive numbers of private vehicles have led to traffic congestion and car accidents, which gradually become a new set of challenges that every modern city must confront [1]. According to the Global Status Report on Road Safety released by the World Health Organization (W.H.O.), over 50 million people get injuries, and 1.3 million individuals lose their lives in car accidents each year worldwide [2]. In the United States alone, over 2.3 million people are injured, and around 40,000 lives are lost in car accidents [3]. Among these car accidents, around 75% are attributed to human errors, such as drowsy driving, driving under the influence (DUI), and distracted driving. The Automated Driving System (ADS) benefits from powerful and robust autonomous driving algorithms, which has the potential to significantly reduce car accidents caused by human mistakes [4], thereby becoming a viable solution to these urgent traffic challenges [5]. The Society of Automotive Engineers (SAE International) categorizes autonomous vehicles into six levels, ranging from Level 0 (fully manual driving) to Level 5 (fully autonomous driving) [6]. Particularly, vehicles at SAE Levels 4 and 5 have the capability to dramatically decrease accidents caused by human mistakes since the algorithm has robust and steady performance in all traffic conditions [7]. In order to increase the autonomous level, extensive research and testing have been conducted in the field in recent years [8], [9]. However, alongside these developments, new challenges and potential issues have also emerged. Despite the significant advances in high-level planning algorithms [10], which are now capable of generating collision-free and efficient paths even in complex traffic scenarios [11], [12], the performance of low-level control systems sometimes remains unsatisfactory.

To address these issues, plenty of research has already been conducted in this field to develop a robust path tracking controller [13]. The optimization-based control approach is a well-known and widely adopted in autonomous driving development [14]. The path-tracking problem is formulated as an optimization problem, where an objective function is designed to minimize tracking errors while satisfying system dynamics and safety constraints. The Control Lyapunov Function - Control Barrier Function - Quadratic Programming (CLF-CBF-QP) approach has also gained significant attention due to its ability to balance safety and stability in an optimization framework [15], [16], [17], [18], [19]. Similarly, Model Predictive Control (MPC) has also become a popular technique, as it provides an optimization-based closed-loop control solution that integrates both planning and control in real time [20], [21], [22].

Beyond optimization-based methods, various other traditional control approaches such as pure-pursuit controller [23], [24], Stanley controller [25], disturbance observer (DOB) [26], and parameter-space multi-objective PID control [27] have been explored to improve ensure accurate and reliable path tracking. While these conventional control theories may perform well in simulation environments, their performance often deteriorates when implemented on real vehicles. This is largely due to the discrepancies between the vehicle model and the actual vehicle dynamics and the presence of inevitable communication delays and computation delays in connected autonomous vehicle (CAV) systems. Such delays can lead to degraded tracking accuracy, oscillatory responses, or even instability, especially when vehicles operate under dynamic and uncertain traffic conditions. Traditional control strategies, such as Proportional–Integral–Derivative (PID) controllers or standard disturbance observer (DOB) approaches, often struggle to maintain high tracking performance under these delay conditions. Therefore, there is a pressing need for an effective and delay-tolerant control method that can ensure precise path tracking and robust stability even when undesirable time delays are present. To address this challenge, this study proposes a Communication Disturbance Observer (CDOB)-based delay-tolerant control framework that compensates for the adverse effects of time delays. In this framework, time-delay is considered as a special kind of disturbance and CDOB is utilized to estimate and compensate for this delay-induced disturbance, which can effectively restore the system's behavior to its delay-free equivalent. After the CDOB compensates for the delay, conventional robust control theories can be directly applied to achieve robust and accurate path tracking, even in the presence of unknown and varying delays. In this paper, parameter-space-designed PID control is used as tracking controller with CDOB since it exhibits good robustness and tracking performance under delay-free conditions. The contributions of this paper are as follows:

- In this paper, we introduce a novel delay-tolerant CDOB framework that models time delays as equivalent disturbances and actively rejects them through modified DOB design, enabling delay-free equivalent dynamics.
- In this paper, we demonstrate the compatibility of the CDOB framework with traditional controllers, allowing robust designs such as parameter-space PID to be seamlessly integrated.
- In this paper, we conduct quantitative evaluations under diverse delay scenarios, showing that the proposed method maintains high tracking accuracy and robustness where conventional PID control response deteriorates.

The remainder of this paper is organized as follows. Section 2 presents the methodologies applied in this study, including the linear path-tracking model, the formulation of the CDOB compensator, and the design strategy of PID tracking controller used in CDOB-based control framework. Section 3 demonstrates the simulation results of the proposed framework, which include path-tracking performance using standard DOB and CDOB controller under various traffic conditions. Finally, Section 4 concludes the paper and outlines possible directions for future research.

## 2. Methodology

To develop a high-performance path tracking controller for CAVs, we propose a modified CDOB based control framework which models time delays as equivalent disturbances and actively rejects them. While the CDOB architecture is able to move the time delay out of the feedback loop for reference following applications, it fails in doing this for disturbance rejection. Since the road curvature enters the feedback loop as a disturbance to be rejected for autonomous vehicle path following, CDOB is not able to compensate for time delays within the loop. A significant contribution of this paper is the development of a modified CDOB architecture that can move the time delay outside the feedback loop in the presence of a disturbance which is typical in autonomous vehicle path following.

First, we present path generation and a linear path-tracking model which is also known as the bicycle model to describe the lateral motion of the ego vehicle, which serves as the foundation for later controller design. Second, we intentionally introduce a time delay into this linear path-tracking model and design a modified CDOB compensator to estimate and reject the delay-induced disturbance, effectively restoring the system's delay-free behavior. It should be noted that such time delay naturally exists in real life CAV path tracking control due to sensor perception system delays, computational delays, drive-by-wire CAN (Control Area Network) bus delays and actuator delays. In order to show the effects of such delays on performance if they are not accounted for, we design a parameter-space-based PID path-tracking controller under delay-free conditions as a benchmarking system. The modified CDOB compensator aims to take the time delay outside the control loop, hence guaranteeing the designed stability and performance albeit after a time delay. The PID benchmarking controller is then utilized in combination with the modified CDOB compensator to achieve robust and precise path tracking and show this built-in delay tolerance, and independently by itself to serve as a baseline for performance comparison if delay compensation is not used. The proposed modified CDOB system can be used in other sensors applications to compensate for the effect of sensor delays on overall feedback control system performance.

### 2.1. Path Generation

Path generation is critical for controller testing, as it provides a realistic reference for evaluating the controller's tracking performance under conditions that closely resemble real-world driving scenarios. The overall procedure for obtaining such a reference path can be summarized as follows: (a) generate a limited number of sample waypoints to represent the general shape of the path; (b) create dense waypoints based on these samples to complete the path design; (c) apply segmentation to the dense waypoints to divide the path into several segments, ideally ensuring that each segment contains a minimal number of features (e.g., corners); and (d) perform polynomial-fit optimization to derive a smooth path expression that ensures continuous curvature within each segment and smooth transitions between adjacent segments; The detailed steps of this procedure are outlined in [28]. Through extensive observations, we find that real vehicles frequently perform lane-change maneuvers while driving as examples of path tracking. To reflect this behavior, we designed two commonly used lane-change patterns: a single-lane-change maneuver and a double-lane-change maneuver. Both patterns were incorporated into the path generation process to create realistic reference trajectories, enabling more comprehensive evaluation of the controller's performance under typical driving scenarios. Figure 1 shows the optimized reference path and its path curvature for the single-lane change maneuver, demonstrating the smoothness of such a path generated using this approach. Similarly, Figure 2 demonstrates the optimized reference path and its path curvature for the double-lane change maneuver

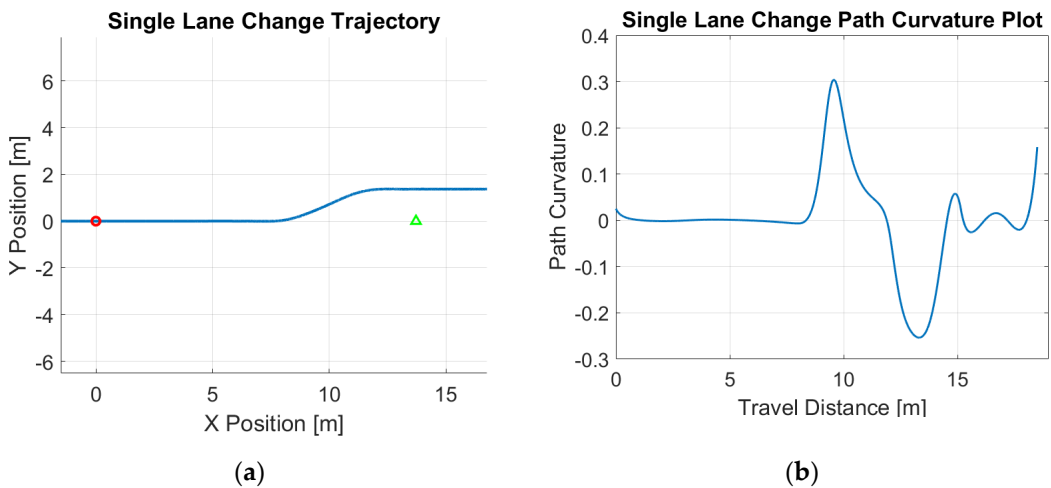

(a)　　　　　　　　　　　　　　(b)

**Figure 1**. Single lane change reference path: (a) trajectory; (b) path curvature

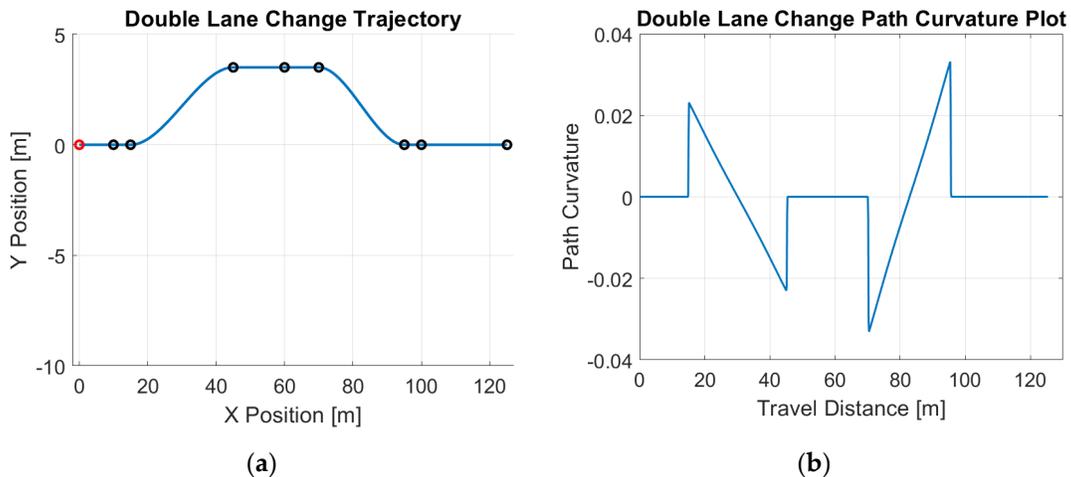

(a)　　　　　　　　　　　　　　(b)

**Figure 2**. Double lane change reference path: (a) trajectory; (b) path curvature

In addition, path-tracking controllers are often integrated with collision avoidance algorithms to ensure both accuracy and safety in autonomous driving systems. To capture this scenario, we employed the well-established traditional collision avoidance method, the Elastic Band algorithm [11], [12], [27], to generate a collision-free trajectory. This trajectory was then used as a reference path for the path-tracking controller, allowing us to evaluate its ability to accurately follow dynamically adjusted paths. Figure 3 shows the reference path and its corresponding path curvature for the collision-avoidance scenario. In Figure 3, the straight path from zero to close to 100 m in the x direction is modified by the Elastic Band algorithm between 20 and 50 m in the x direction to avoid an obstacle.

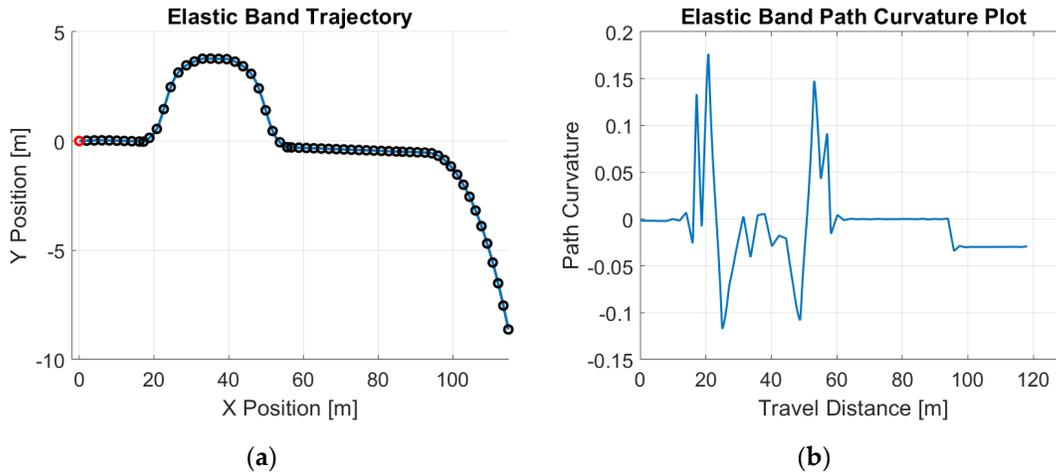

(a)      (b)

**Figure 3**. Elastic-Band generated reference path: (a) trajectory; (b) path curvature

2.2. Linear Path-tracking Model

This section presents the linear path-tracking model that serves as the basis for the proposed control routine. The detailed derivation of this model can be found in [28] and [29]. Similar models that have been derived using a similar approach have also been applied to articulated vehicle configurations in the literature [30]. This linear path-tracking model contains two components: a linear lateral single-track model and a path-tracking model augmentation. The plane's lateral motion of this single-track vehicle is illustrated in Figure 4, and the resulting linear single-track vehicle dynamic model is described in Equation (1). The overall model is given in Equations (2) through (5) with an explanation of the parameters used being presented in Table 1.

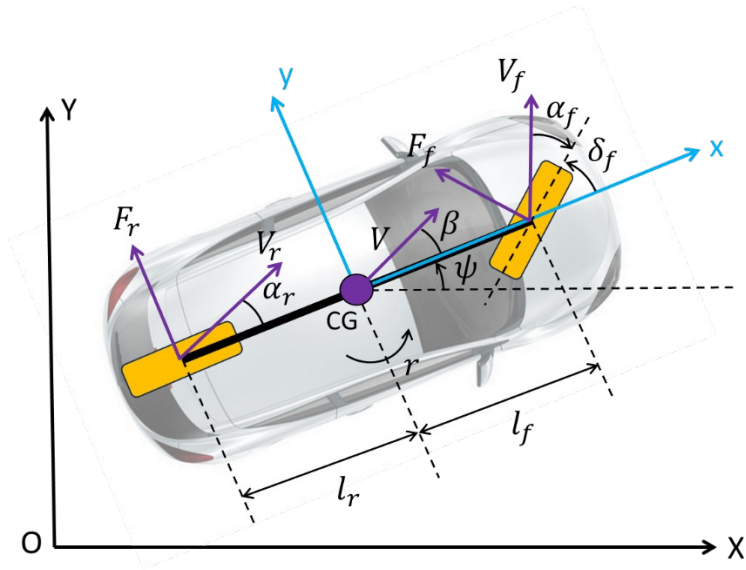

**Figure 4.** Linear single-track lateral vehicle dynamic model.

$$\begin{bmatrix}\dot{\beta}\\\dot{r}\end{bmatrix} = \begin{bmatrix}\dfrac{-C_f - C_r}{mV} & -1 + \left(\dfrac{C_r l_r - C_f l_f}{mV^2}\right)\\ \dfrac{C_r l_r - C_f l_f}{I_z} & \dfrac{-C_f l_f^2 - C_r l_r^2}{I_z V}\end{bmatrix}\begin{bmatrix}\beta\\r\end{bmatrix} + \begin{bmatrix}\dfrac{C_f}{MV} & \dfrac{C_r}{MV}\\ \dfrac{C_f l_f}{I_z} & \dfrac{C_r l_r}{I_z}\end{bmatrix}\begin{bmatrix}\delta_f\\\delta_r\end{bmatrix} + \begin{bmatrix}0\\\dfrac{1}{I_z}\end{bmatrix}M_z \quad (1)$$

$$\Delta\dot{\Psi}_p \cong r - V\rho_{ref} \quad (2)$$

$$\dot{e}_y \cong V\beta + l_s r + V\Delta\Psi_p - l_s V\rho_{ref} \quad (3)$$

$$\rho_{ref} = \dfrac{\dfrac{dx_p(\lambda)}{d\lambda}\dfrac{d^2 y_p(\lambda)}{d\lambda^2} - \dfrac{dy_p(\lambda)}{d\lambda}\dfrac{d^2 x_p(\lambda)}{d\lambda^2}}{\left(\left(\dfrac{dx_p(\lambda)}{d\lambda}\right)^2 + \left(\dfrac{dy_p(\lambda)}{d\lambda}\right)^2\right)^{\frac{3}{2}}} \quad (4)$$

$$\begin{bmatrix} \dot{\beta} \\ \dot{r} \\ \Delta\dot{\psi}_p \\ \dot{e}_y \end{bmatrix} = \begin{bmatrix} \dfrac{-C_f - C_r}{MV} & -1 + \dfrac{C_r l_r - C_f l_f}{MV^2} & 0 & 0 \\ \dfrac{C_r l_r - C_f l_f}{I_z} & \dfrac{-C_f l_f^2 - C_r l_r^2}{I_z V} & 0 & 0 \\ 0 & 1 & 0 & 0 \\ V & l_s & V & 0 \end{bmatrix} \begin{bmatrix} \beta \\ r \\ \Delta\psi_p \\ e_y \end{bmatrix}$$

$$+ \begin{bmatrix} \dfrac{C_f}{MV} & \dfrac{C_r}{MV} \\ \dfrac{C_f l_f}{I_z} & \dfrac{C_r l_r}{I_z} \\ 0 & 0 \\ 0 & 0 \end{bmatrix} \begin{bmatrix} \delta_f \\ \delta_r \end{bmatrix} + \begin{bmatrix} 0 \\ 0 \\ -V \\ -l_s V \end{bmatrix} \rho_{ref} + \begin{bmatrix} 0 \\ \dfrac{1}{I_z} \\ 0 \\ 0 \end{bmatrix} M_{zd} \quad (5)$$

**Table 1**. Linear path-tracking model parameters [18].

| Parameter | Explanation |
|---|---|
| $\beta$ | Vehicle side slip angle |
| $r$ | Vehicle yaw rate |
| $\Delta\psi_p$ | Heading error |
| $e_y$ | Path-tracking error |
| $C_f$ | Front tire cornering stiffness (195,000 $N/rad$) |
| $l_f$ | Distance between CG and front axle (1.3008 $m$) |
| $C_r$ | Rear tire cornering stiffness (50,000 $N/rad$) |
| $l_r$ | Distance between CG and rear axle (1.5453 $m$) |
| $M$ | Vehicle mass (1997.6 $kg$) |
| $V$ | Vehicle velocity |
| $l_s$ | Preview distance |
| $I_z$ | Vehicle yaw moment of inertia (3728 $kg * m^2$) |
| $\rho_{ref}$ | Reference path curvature |
| $M_{zd}$ | Yaw moment disturbance |
| $K$ | Preview distance scheduling constant |
| $t_d$ | Time Delay (0.01/0.05/0.1/0.3 $s$) |

It can be observed that, for generality, the model presented in Equation (1) has both front and rear wheel steering angles $\delta_f$ and $\delta_r$ as inputs. In our case, the vehicle is assumed to be front-wheel-steer only. To effectively employ this vehicle lateral dynamic model for path tracking, path tracking heading error $\Delta \Psi_p$ and path tracking error $e_y$, which contain both deviation error and heading error, are introduced. Equations (2) and (3) demonstrate how to calculate the path tracking heading error $\Delta \Psi_p$ and path tracking error $e_y$. Also, since the path tracking error $e_y$ also depends on the road curvature, Equation (4) demonstrates how to calculate road curvature based on road shape. $x_p$ and $y_p$ are coordinates of the optimized path. Equation (5) is the state space equation for the combined linear path-tracking model. It can also be noticed that path curvature $\rho_{ref}$ and yaw moment disturbance $M_{zd}$ enter the model as external disturbances. In practice, the yaw moment disturbance is often ignored as it is usually considered in a separate yaw stability control system [8], [26], [31]. Additionally, the preview distance $l_s$ is chosen to be a linear function of vehicle speed. It should also be noted that vehicle speed can be scheduled according to the refence path curvature to make sure vehicle lateral acceleration stays within an acceptable limit. The parameter values used in the simulations are also listed in parentheses in Table 1.

### 2.3. Modified Communication Disturbance Observer Design

#### 2.3.1. Modified CDOB Framework

This section presents a general overview of the modified communication disturbance observer (CDOB), which is an approach inspired by the disturbance observer (DOB). Please see references [27], [32], [33] for more details on the CDOB. Given a time-delayed input-output system as shown in Figure 5a, an equivalent system can be constructed as displayed in Figure 5b, where a term, $D(s)$, incorporates the time delay and is fed into the system as a disturbance. It must be noted that the value of the time delay is not necessarily known, which makes the CDOB a very useful method for practical applications.

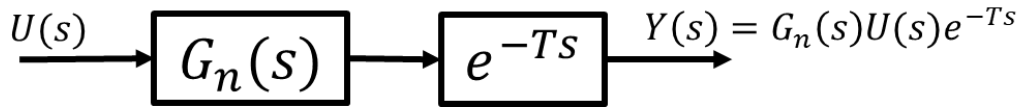

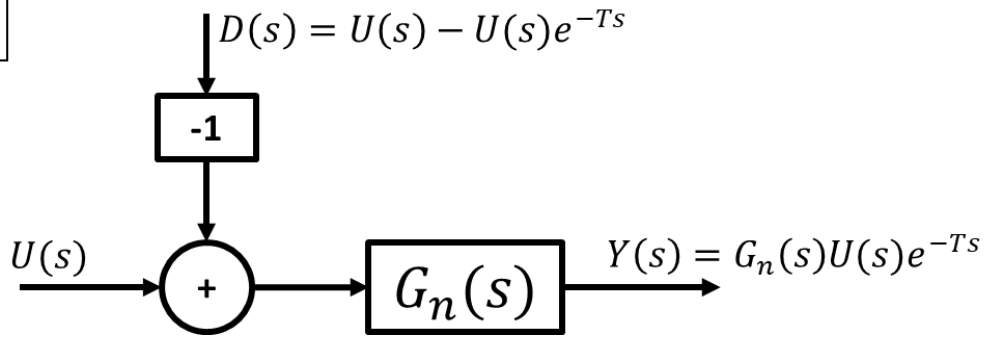

**Figure 5.** Equivalent representations of sample input-output system with time delay.

Once it has been established that the unknown time delay can be remodeled as a disturbance, the concept of DOB can be used to estimate and compensate for the time delay. Figure 6 shows the basic structure of the CDOB consisting of a time delay estimation loop and a time delay compensation loop. $Q(s)$ in the time delay estimation loop is a unity-gain low-pass filter of the appropriate order introduced to ensure that $Q(s)/G_n(s)$ is proper, hence ensuring that the scheme is implementable. Assuming that the analysis is carried out at low frequency where $Q(s) = 1$, it can be derived that the output of the time delay estimation loop yields $\bar{D}(s)$ which is an estimation of $D(s)$ as shown in Figure 6. The additional time delay compensation loop cancels out the term containing time delay and yields the desired output form $G_n(s)U(s)$ that is not affected by the unknown time delay.

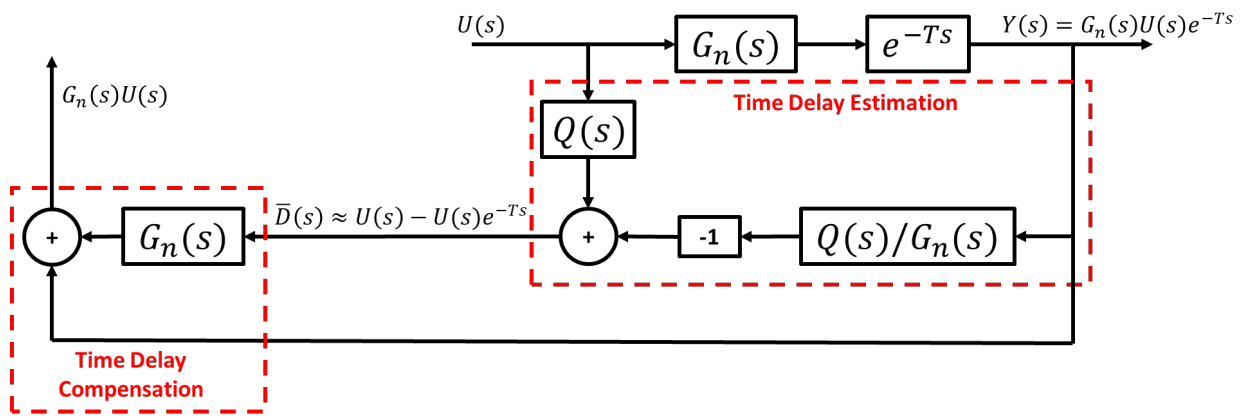

**Figure 6.** Standard CDOB block diagram.

In this paper, the Butterworth approach is utilized to design the unity-gain low-pass filter $Q(s)$ in the CDOB loop. Table 2 displays the parameters used in the design. Equations (6) and (7) demonstrate how to calculate the order and cut-off frequency of the proposed low pass filter $Q(s)$. From the results, we notice that the minimum order of this low pass filter is second. Then, these results are substituted into the second order Butterworth low pass filter equation which results in Equation (8).

**Table 2**. Low Pass Filter $Q$ Butterworth Approach Parameters

| Parameter | Explanation | Value |
|---|---|---|
| $\omega_p$ | Passband Frequency | $1000 \ rad/s$ |
| $\omega_s$ | Stopband Frequency | $10000 \ rad/s$ |
| $\alpha_p$ | Maximum Passband Attenuation | $3 \ dB$ |
| $\alpha_s$ | Maximum Stopband Attenuation | $30 \ dB$ |

$$N \geq \frac{\log \frac{10^{\alpha_s/10} - 1}{10^{\alpha_p/10} - 1}}{2 * \log \frac{\omega_s}{\omega_p}} = 1.5008 \quad (6)$$

$$\Omega_{cutoff} = \frac{\omega_p}{(10^{\alpha_p/10} - 1)^{\frac{1}{2N}}} = 1001.6 \frac{rad}{s} \quad (7)$$

$$Q = \frac{1}{\frac{s}{\Omega_c}^2 + 1.4142 \frac{s}{\Omega_c} + 1} \quad (8)$$

One final remark is that the CDOB can effectively restore the system's behavior to its delay-free equivalent and usually needs to be applied together with an additional feedback controller. This feedback controller is used to take system references $R(s)$ and delay-free feedback $G_n(s)U(s)$ as input and generate control signal $U(s)$.

The feedback controller mentioned above can be of any design. We present an example design that features a speed-scheduled, parameter-space PID controller, the details of which can be found in [28]. The parameter-space method is discussed in detail in [27], also in references [34], [35], [36] that focus on application to autonomous path following. The form of the controller and its detailed design are presented in the following subsections.

## 2.3.2. Modification for Path Curvature Rejection

As mentioned in Section 2.2, the vehicle model used for path-tracking is derived such that reference path curvature enters the model as an external disturbance. Denoting the path curvature disturbance as $d$, the desired output of the CDOB hence becomes $G_n(s)U(s) + d$. Adding this disturbance $d$ into the CDOB block diagram as illustrated in Figure 7, however, does not yield the desired outcome, where the actual output remains in the form of $G_n(s)U(s)$, lacking the disturbance term $d$.

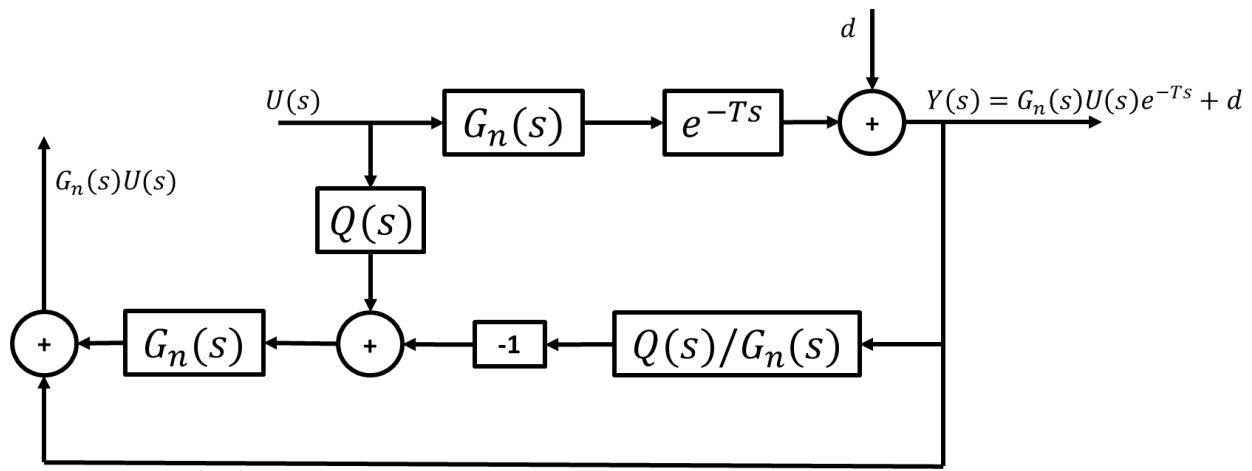

**Figure 7**. Standard CDOB block diagram with external disturbance.

To account for the above issue, modifications must be made to the standard CDOB structure to accommodate the path curvature rejection requirement. Figure 8 shows the modified CDOB block diagram, where the same path curvature disturbance is added to the output of the CDOB delay compensation loop. It should be noted that this structure works because the curvature of the reference path, i.e. the disturbance in CAV path following, is known.

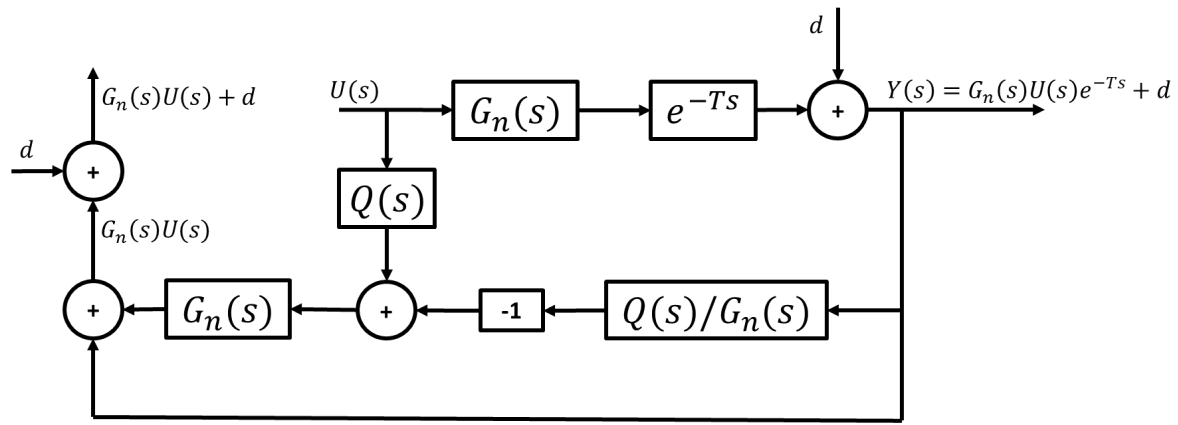

**Figure 8**. Modified CDOB block diagram with external disturbance.

## 2.3.3. Feedback Controller Design

As mentioned in section 2.3.1, with the modified CDOB capable of outputting desired output form without the interference of time delay, a closed-loop control system is required to be constructed for this non-time-delayed disturbance rejection problem. In Figure 9, a generic feedback controller $C(s)$ is added to generate input $U(s)$ such that reference input $R(s)$ can be tracked.

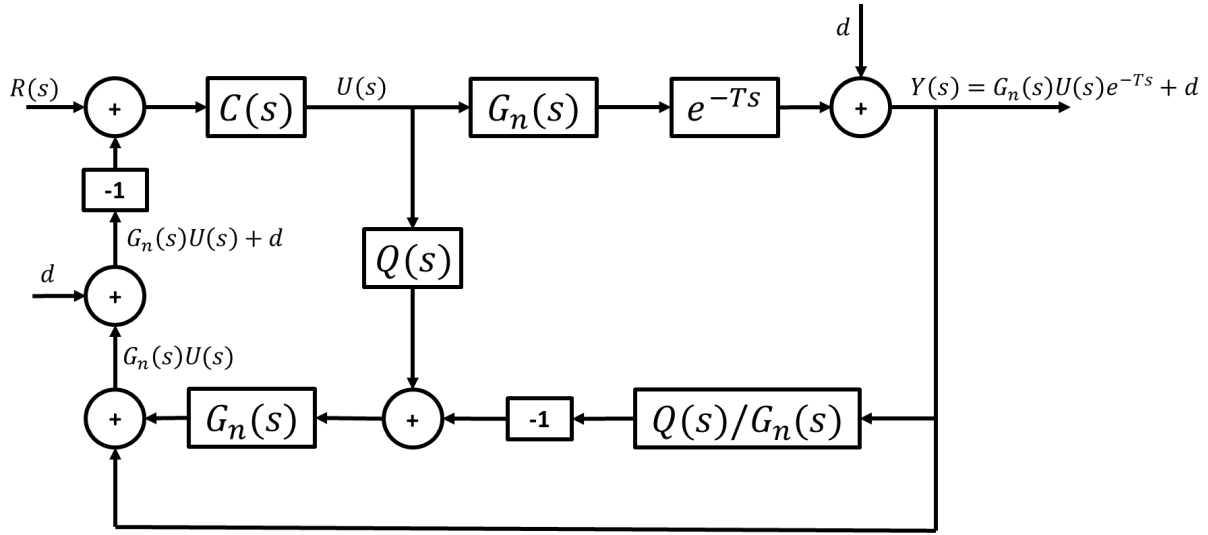

**Figure 9**. Control system with modified CDOB and feedback controller

In this paper, a parameter-space PID controller is designed for precise and robust path following as C(s) in Fig. 9. The reference input $R(s)$ should be zero in this case noting that the goal of the control system is to eliminate path-tracking error $e_y$. The form of the controller is presented in Equation (9).

$$C(s) = k_p + \frac{k_i}{s} + k_d s \qquad (9)$$

The controller gains $(k_p, k_i, k_d)$ are the parameters to be tuned. Since the controller is speed-scheduled, the tunable parameter set has four elements: $(V, k_p, k_i, k_d)$. A D-stability region, as displayed in Figure 10, is established for desired pole placement of the closed loop system. An example of the admissible controller gain region at a certain scheduled speed is shown in Figure 11. It should be remarked that during the process of controller gains value selection, a general rule of thumb is to choose the gains to be as small as possible within the admissible region so that the control effort can be minimized.

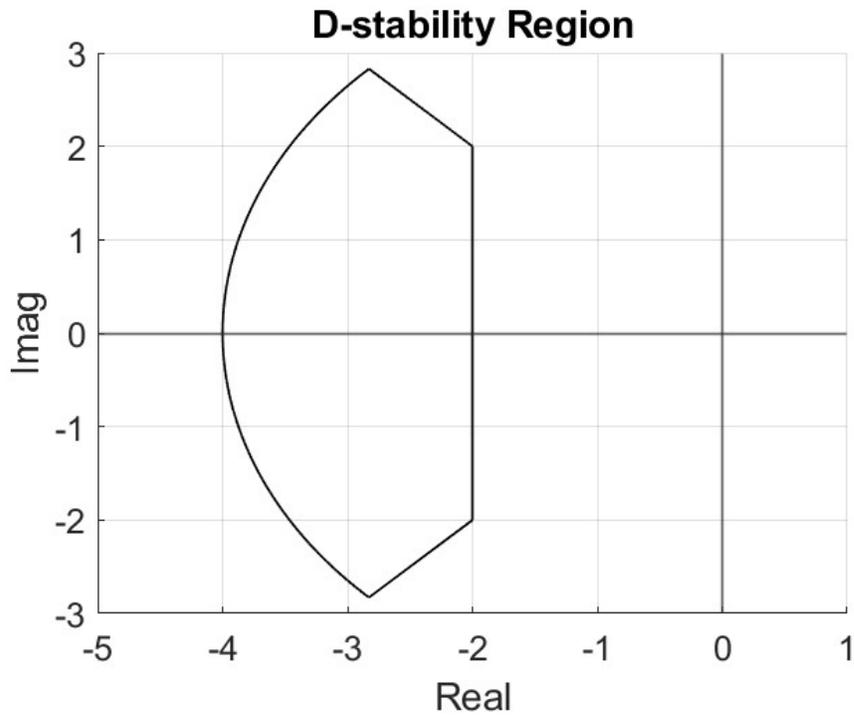

**Figure 10**. D-stability region

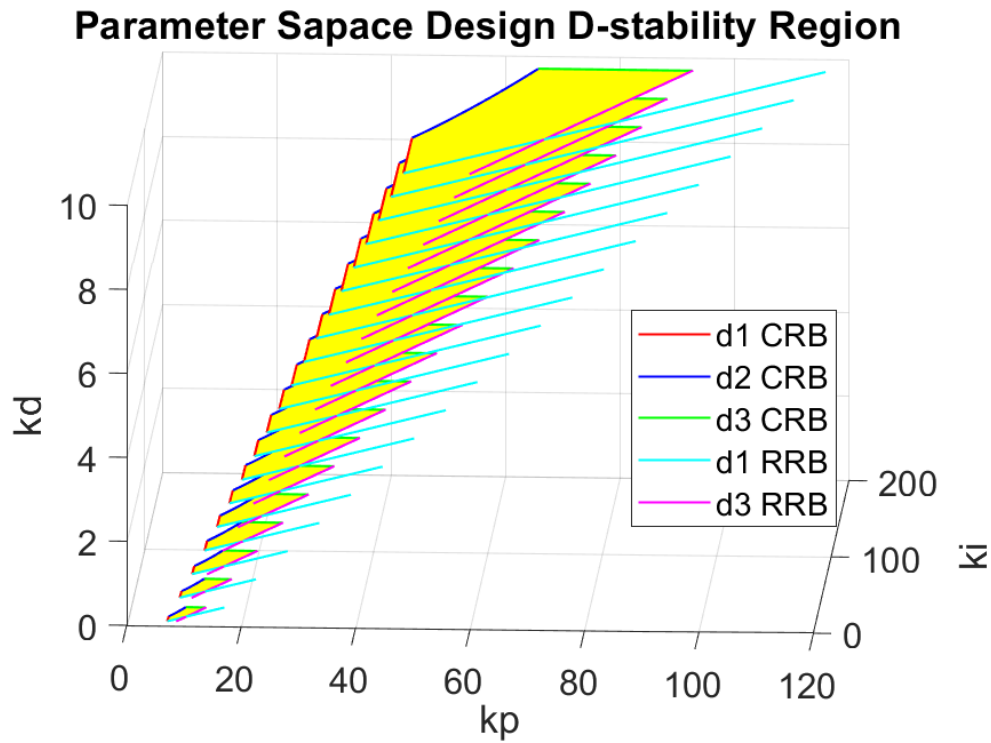

**Figure 11**. Admissible control region at a certain speed

## 3. Results

3.1. Simulation Study

Simulation studies are first performed to demonstrate the efficacy of the proposed control design. A Simulink model is constructed to simulate the motions of the vehicle. The parameter values used in the simulations are listed in Table 1. The example reference path shown in Figures 1, 2 and 3 are used as the desired path in this experiment.

The single lane change motion simulation results for the combined modified CDOB with PID control system and the PID only cases are displayed in Figures 12 and 13. It can be observed that the vehicle is able to track the reference path satisfactorily with reasonably small path-tracking errors by applying smooth steering inputs under the control of CDOB with PID. However, under PID only conditions, the tracking operation will fail in a very early stage without the CDOB feature even only with 0.01s delay. This is due to the detrimental effect of time delay on system stability. The PID no delay case which is the desired result is shown in Figure 12b. The PID+CDOB compensated path following responses for different time delay values displayed in Figure 12a are all very similar to this ideal PID with no delay response showing the efficacy of the proposed system. The PID results with different delay values in Figure 12b all end with the vehicle running off the path at up to 2 m in the X direction. The path tracking errors shown in Figure 13 demonstrate this more clearly. Figure 13a shows that the PID+CDOB system can follow the path with relatively small error even though error values increase with increases in time delay. The PID only results in Figure 13b show that the path tracking error becomes oscillatory and unstable, showing the need for time delay compensation.

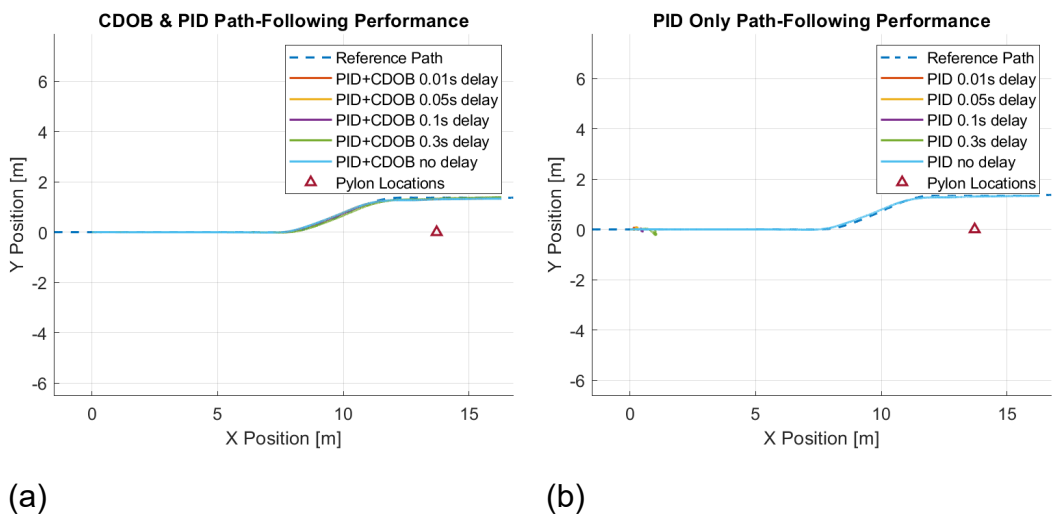

(a)          (b)

**Figure 12.** Single-lane change maneuver simulation results: (a) CDOB+PID; (b) PID

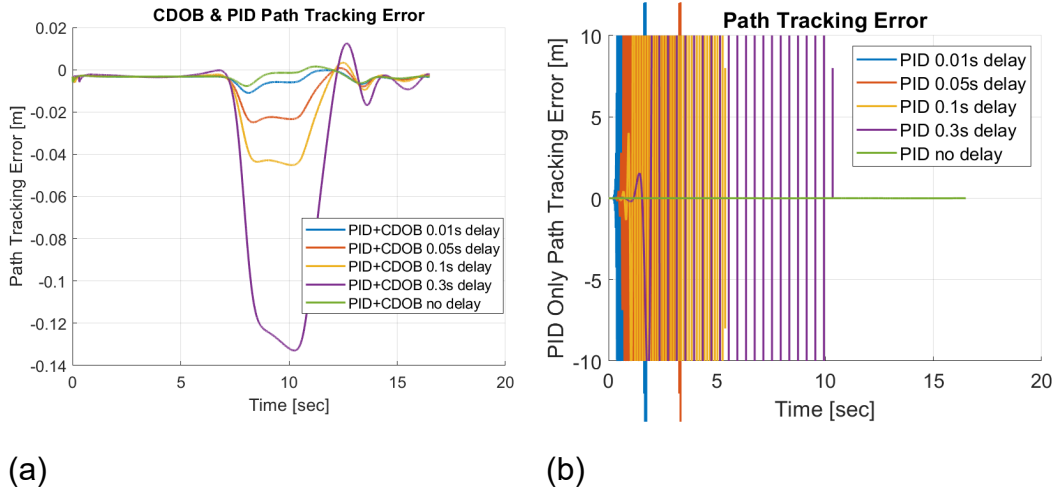

(a)                                 (b)

**Figure 13**. Single-lane change maneuver tracking error: (a) CDOB+PID; (b) PID

Similarly, we perform the same test using the double lane change trajectory shown in Figure 2. The double lane change motion simulation results for the combined modified CDOB and PID control system and PID only are displayed in Figures 14 and 15. Just like the single-lane change tracking results, the proposed the CDOB+PID framework closely follows the reference path even with delays up to 0.3 s, whereas the PID-only controller exhibits significant trajectory deviations and eventually fails to track the path even with 0.01 s time delay. This is further reflected in the path tracking error plots in Figure 15, where the CDOB+PID controller maintains a small and smooth error profile within ±0.08 m, indicating stable performance. In contrast, the PID-only controller experiences a rapid oscillatory in tracking error, which indicates a loss of stability. These observations highlight that the CDOB framework can effectively compensate for delay-induced disturbances

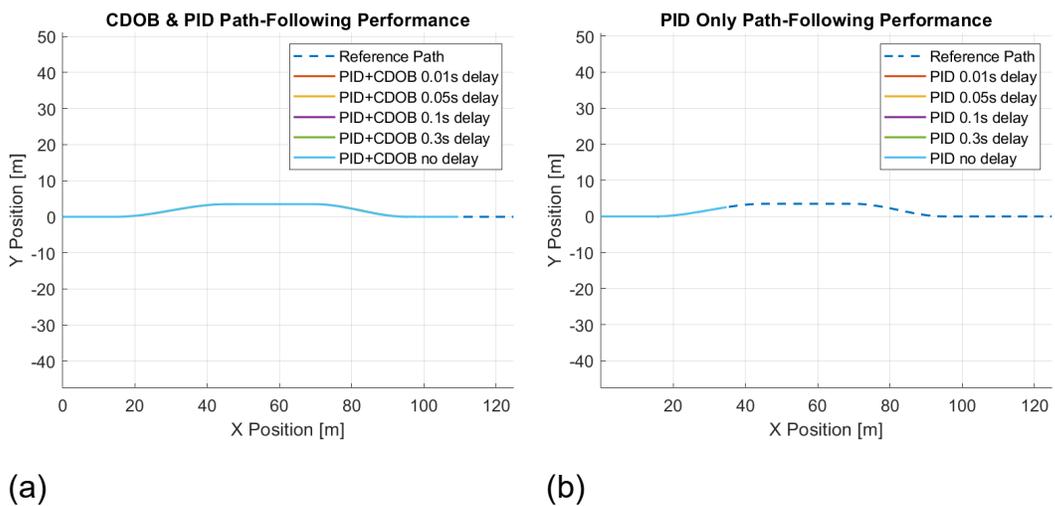

(a)                                 (b)

**Figure 14**. Double-lane change maneuver simulation results: (a) CDOB+PID; (b) PID

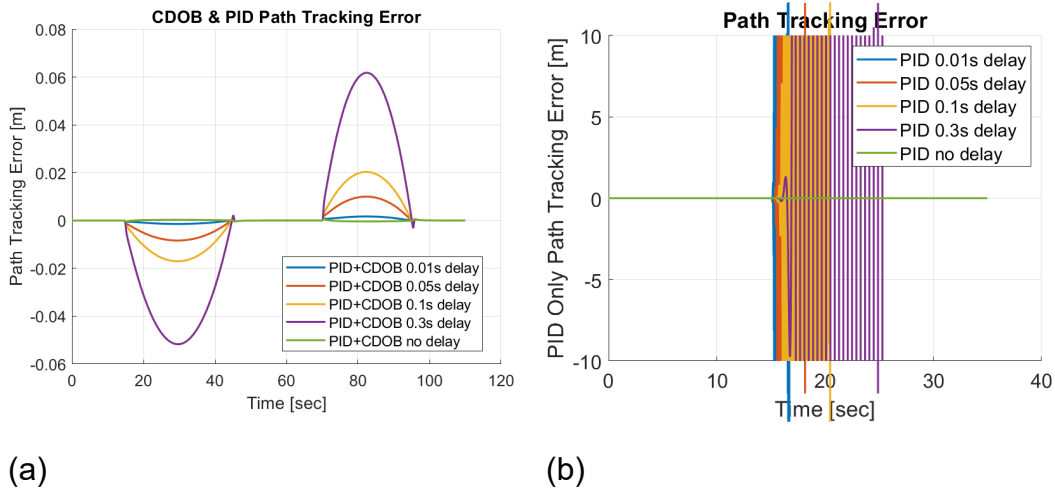

**Figure 15**. Double-lane change maneuver tracking error: (a) CDOB+PID; (b) PID

In order to evaluate the performance of proposed CDOB control framework within an autonomous driving system, the Elastic Band algorithm is used to generate a collision-free trajectory as shown in Fig. 3. A similar test is performed using this trajectory and similar tracking results can be observed, further demonstrating the effectiveness of the CDOB+PID framework under time delay conditions. As shown in Figure 16, the CDOB+PID controller accurately follows the reference trajectory across all tested delay levels, with only minimal deviations even at a 0.3 s delay. In contrast, the PID-only controller again suffers from time delays and cannot successfully track the reference path. Path tracking error plots shown in figure 17 demonstrate that the modified CDOB+PID controller can bound errors within approximately ±0.2 m while the PID-only controller leads to system instability under time delay. These results confirm that the modified CDOB framework effectively mitigates delay-induced disturbances, ensuring accurate path following and stability not only for standard maneuvers but also for dynamically generated trajectories such as those from the elastic band or other collision avoidance algorithms.

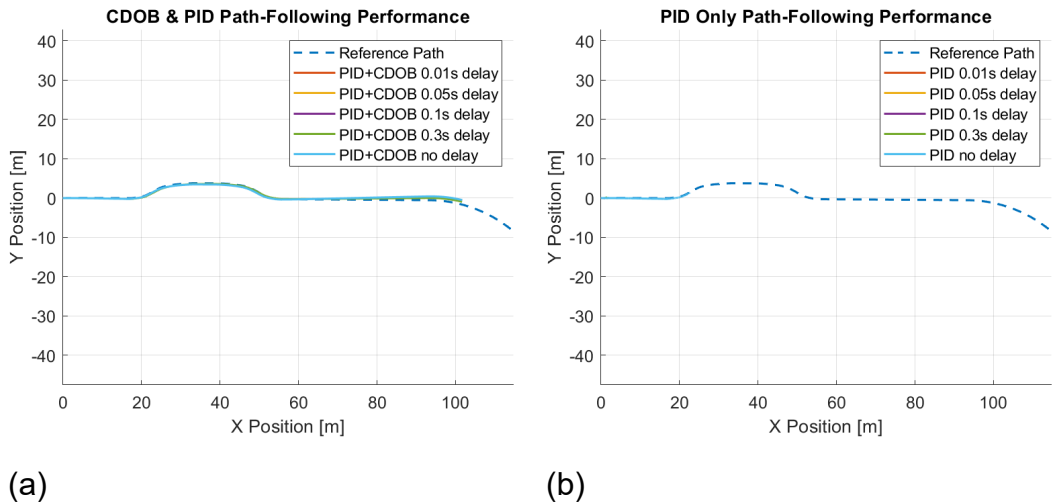

**Figure 16**. Elastic Band collision avoidance maneuver simulation results: (a) CDOB+PID; (b) PID

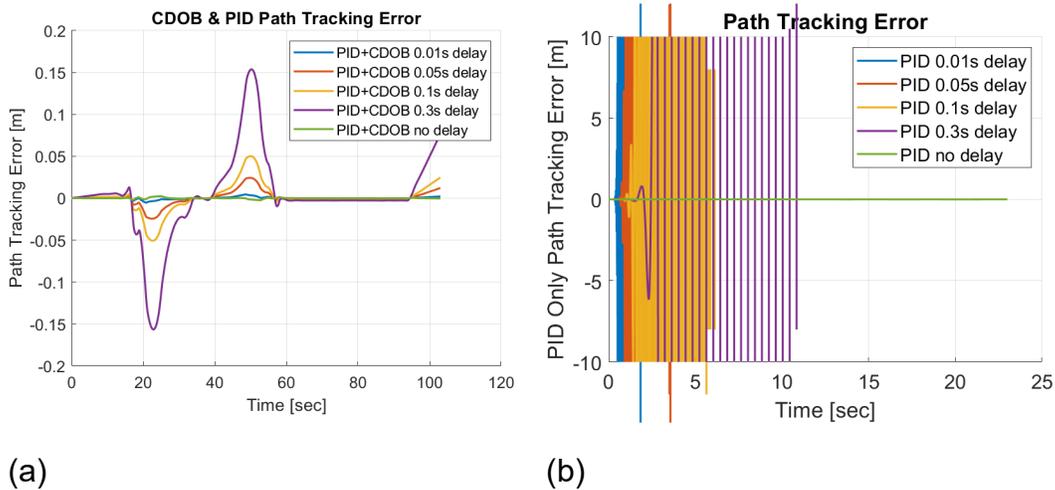

**Figure 17**. Elastic Band collision avoidance maneuver tracking error: (a) CDOB+PID; (b) PID

3.2. Hardware-in-the-loop (HIL) Experiments

In addition to the simulation study, hardware-in-the-loop (HIL) experiments are also performed to further demonstrate the suitability of this proposed modified CDOB control framework for real-life implementations. The same parameter choices as shown in Table 1 as well as the same example path are used. For the time-delay condition, a medium delay of 0.1 s was selected, as it reasonably represents typical operating scenarios that include both communication and computational delays. The single-lane change tracking results are shown in Figure 18. It can be observed that even for online operation in the HIL simulator, the proposed control scheme is able to follow the desired path effectively with small path-tracking errors. The maximum error remains below 0.05 m demonstrating high tracking accuracy. The steering input profile remains smooth and well within the saturation limits, highlighting the controller's ability to generate stable and feasible control commands. The HIL results validate that the proposed CDOB control framework not only performs reliably in simulations but also maintains accuracy, stability, and feasibility under real-time hardware constraints, making it highly suitable for practical autonomous driving implementations.

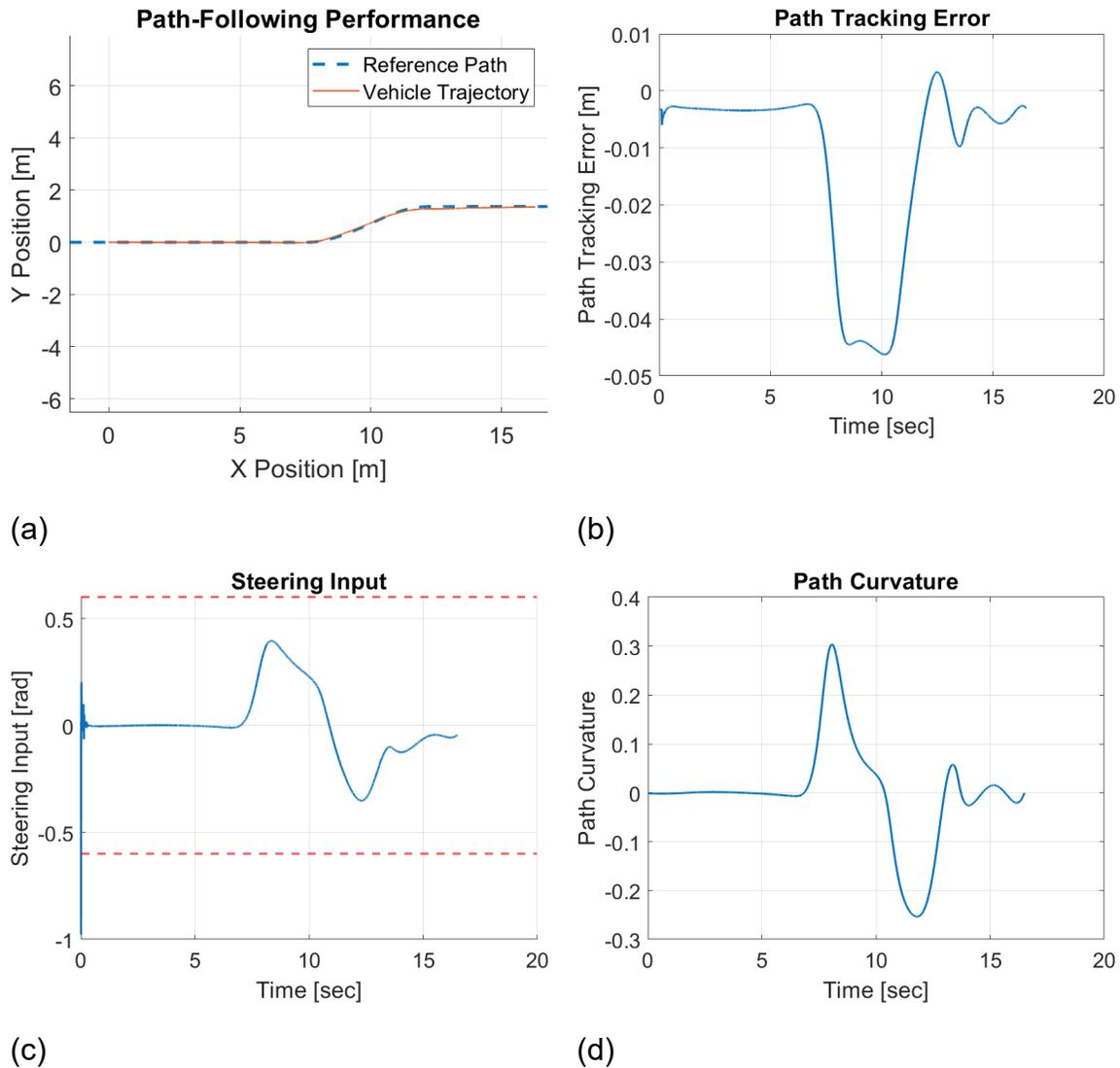

**Figure 18**. Forward motion modified CDOB + PID HIL results

Similar HIL test is performed in double-lane change trajectory. The experiment results in Figure 19 demonstrate that the vehicle trajectory closely matches the reference path across the entire maneuver with only minor deviations under the control of the proposed modified CDOB+PID controller. The path-tracking error remains within ±0.02 m, indicating precise and stable path tracking. It can also be observed that the large path tracking error often occurs during the curved sections of the trajectory, where sudden changes in path curvature leads to more aggressive control inputs, resulting in larger tracking errors. The steering input profile stays smooth and well within the saturation limits, suggesting that the controller generates stable and feasible commands.

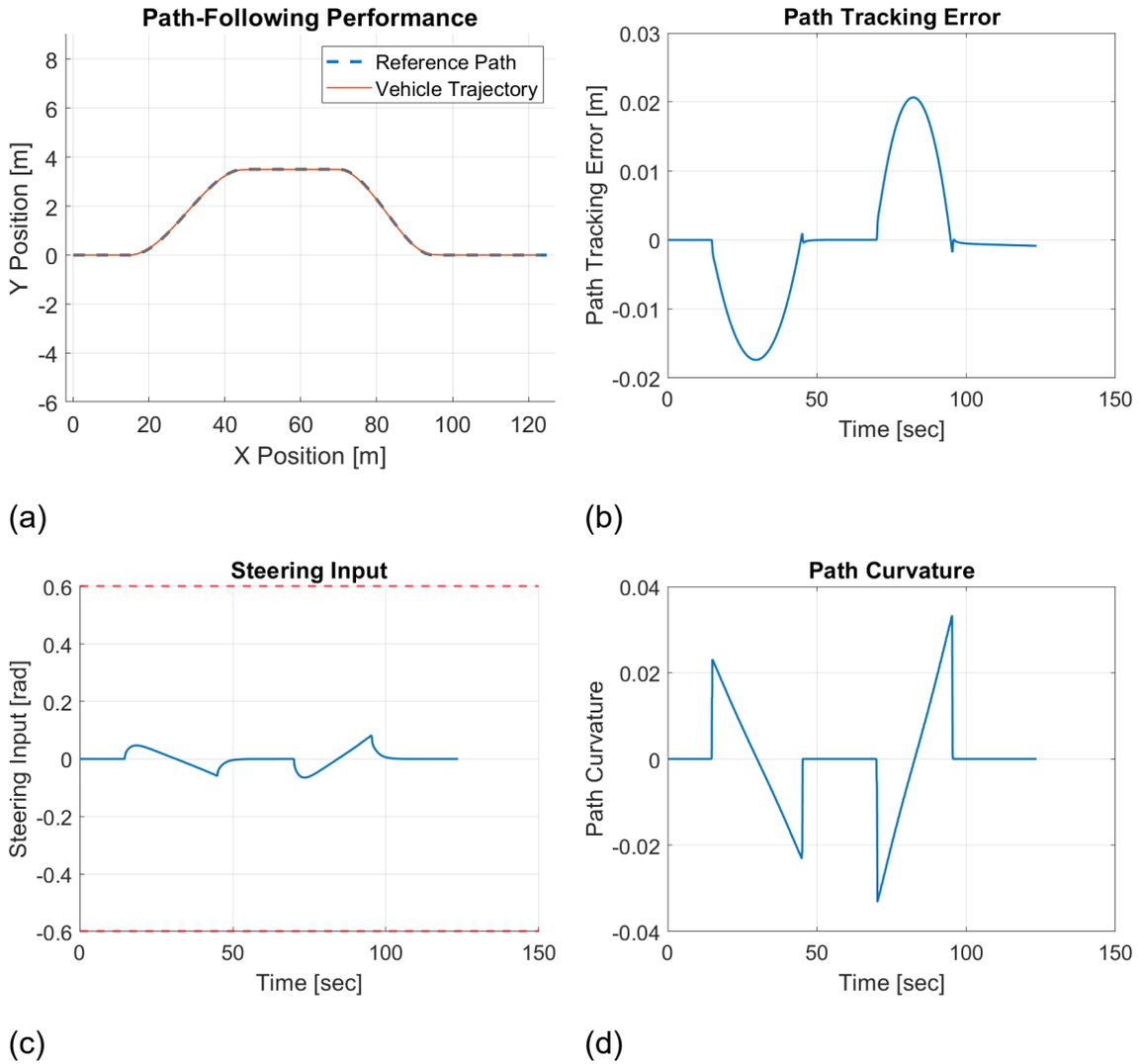

**Figure 19**. Forward motion modified CDOB + PID HIL results

A similar HIL test was performed using the Elastic Band–generated collision-free trajectory to evaluate the performance of the proposed framework in an autonomous driving setting. The experimental results in Figure 21 demonstrate that the vehicle can accurately follow the reference path under the control of the proposed CDOB+PID controller. The path-tracking errors increase primarily in sections with larger path curvature, particularly in the middle collision-avoidance segment and near the endpoint.

Nevertheless, the steering input profile remains smooth and well within the saturation limits.

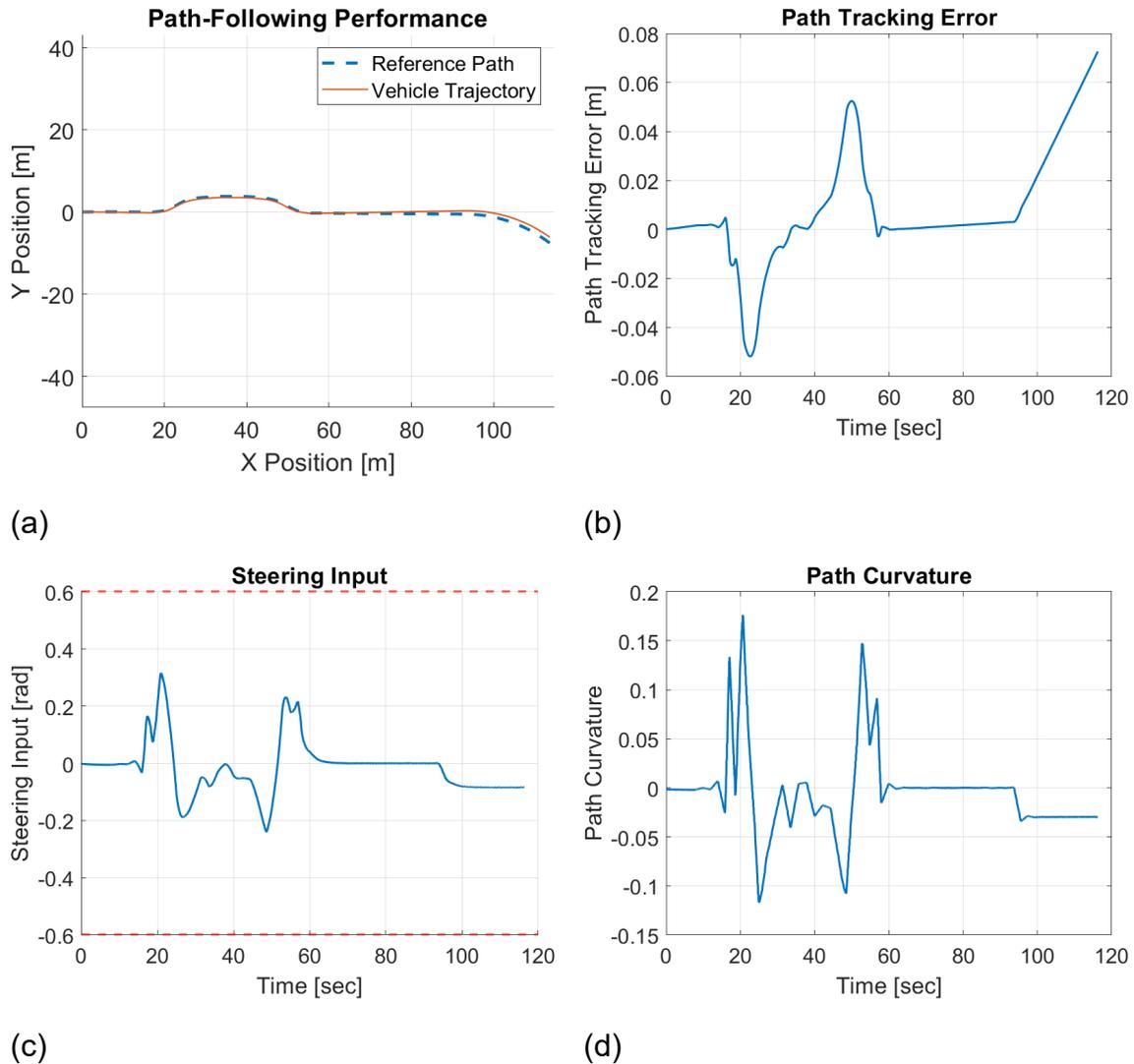

**Figure 20**. Forward motion modified CDOB + PID HIL results

## 4. Conclusions

This paper addressed the critical challenge of communication and computation delays in CAV path-tracking control. Although conventional control methods demonstrated robust performance under delay-free conditions, their effectiveness degraded significantly when subjected to unknown and varied delays. To overcome this limitation, we proposed a delay-tolerant CDOB framework that models time delays as equivalent disturbances and actively compensates for them. By effectively removing the impact of time delays, the CDOB restores the system's behavior to its delay-free equivalent, enabling traditional control theories to operate at their full potential. The simulation and HIL experiment results demonstrate the robustness and accuracy of the proposed CDOB+PID control framework in path tracking control. In case studies, the proposed controller maintained close alignment with the reference trajectory across various scenarios, including single-lane change, double-lane change, and Elastic Band–generated collision-avoidance paths, even under time delays of up to 0.3 s. Path-tracking errors remained small, typically within ±0.08 m for standard maneuvers and ±0.2 m for more complex trajectories, while steering inputs stayed smooth and well within feasible limits. These findings demonstrate that CDOB offers a practical and effective solution for achieving reliable path tracking in real-world CAV applications where delays are inevitable.

Future work will focus on extending the current study in several directions to further explore and enhance the potential of the proposed modified CDOB-based control framework for autonomous driving applications. First, while this study tested only constant time-delay conditions, and the CDOB framework is theoretically capable of compensating for arbitrary delays, further validation under varying and time-varying delay scenarios is required. Second, we plan to integrate more advanced planning and collision-avoidance modules into the testing framework to comprehensively evaluate the controller's performance in complex, dynamic traffic conditions. The current implementation utilizes a parameter-space-designed PID controller as the nominal controller. Future studies will investigate the potential of coupling CDOB with more advanced nominal controllers to fully exploit its capabilities and improve performance in real-world autonomous driving systems. The stability conditions for the modified CDOB can be investigated analytically using methods like the ones in references [37], [38], [39] and will be considered in future work. The method can be more specifically applied to autonomous driving [40] and collision avoidance [41]. [42].